# Phase inversion and collapse of the cross-spectral function

C. W. Nelson, A. Hati and D. A. Howe
National Institute of Standards and Technology
Boulder, CO 80305
Email: craig.nelson@boulder.nist.gov

## Abstract

Cross-spectral analysis is a mathematical tool for extracting the power spectral density of a correlated signal from two time series in the presence of uncorrelated interfering signals. We demonstrate and explain a set of conditions where the detection of the desired signal using cross-spectral fails partially or entirely in the presence of a second uncorrelated signal. Not understanding when and how this effect occurs can lead to dramatic underreporting of the desired signal. Theoretical, simulated and experimental demonstrations of this effect as well as mitigating methods are presented.

## Introduction

The detection of a signal in the presence of interfering noise always presents a challenge. The power spectral density (PSD) of an ergodic and stationary signal can be determined from the cross-spectrum even in the presence of interfering noise. If one can create two reproductions of a desired signal and the interfering noise in each copy is not correlated, the average of the cross-spectrum can be used to estimate the PSD of the desired signal even when the intensity of the interfering noise is dominant.

Cross-spectral analysis has been used in the field of modulation noise metrology for improving the sensitivity of measurements for nearly fifty years [1-5]. The last twenty years has seen extensive use of cross-spectral analysis in the laboratory [6-12] and it has even made its way into commercial test equipment, at first with small companies [8,13] and now with major mainstream test equipment manufacturers. It is generally understood when a desired signal has some level of correlation with an interferer, occasionally some level of cancelation of signals can be observed. We will demonstrate and explain a set of conditions where the detection of the desired signal using cross-spectral collapses partially or entirely in the presence of second uncorrelated interfering signal. This effect was first observed by Ivanov and Walls in 2000 [14, 15] and in this paper we refine the previous discussion by proposing that the failure to correctly detect the measured noise occurs when a special phase condition exists between the signals being presented to the cross-spectrum function. Significantly, this paper's findings may negate many overly optimistic and untrue low-noise results of past measurements. Unfortunately, there is no easy way to determine if a past cross-correlation measurement was affected by the cause given in this paper.

## The cross-power spectral density

The cross-spectrum of two signals x(t) and y(t) is defined as the Fourier transform of the cross-covariance function of x and y. However, from the Wiener–Khinchin theorem, it can be implemented far more practically by the metrologist as:

$$
\begin{aligned}
&\hat{X}(f) = F\{x(t)\} \\
&\hat{Y}(f) = F\{y(t)\} \\
&\hat{S}_{xy}(f) = \frac{1}{T}\left\langle \hat{X}(f)\hat{Y}^*(f)\right\rangle,
\end{aligned}
\tag{1}
$$

where, $X(f)$ and $Y(f)$ are the Fourier transforms of $x(t)$ and $y(t)$. The cross-power spectral density $S_{xy}(f)$ can be determined from the ensemble average of $X(f)$ multiplied by the complex conjugate of $Y(f)$. $T$ is the measurement time normalizing the PSD to 1 Hz. The *caret* '^' indicates complex and '*' indicates complex conjugate. Unlike a normal PSD the cross-PSD is a complex quantity. An excellent detailed description of the cross-spectrum can be found in [16]. In this section, we highlight different scenarios of cross-spectrum analysis.

***Case i***: If we have two signals $x(t)$ and $y(t)$, each comprised of three statistically independent, ergodic and random processes $a(t)$, $b(t)$ and $c(t)$ such that

$$
\begin{aligned}
x(t) &= a(t) + c(t) \\
y(t) &= b(t) + c(t).
\end{aligned}
\tag{2}
$$

We consider $c(t)$ to be the desired signal that we wish to recover, and $a(t)$ and $b(t)$ are the uncorrelated interfering signals. The Fourier transforms of these signals are represented by the corresponding capital variables as

$$
\begin{aligned}
\hat{X}(f) &= \hat{A}(f) + \hat{C}(f) \\
\hat{Y}(f) &= \hat{B}(f) + \hat{C}(f)
\end{aligned}.
\tag{3}
$$

$$
\begin{aligned}
\left|\hat{S}_{xy}(f)\right| &= \tfrac{1}{T}\left|\left\langle \hat{X}(f)\hat{Y}^*(f)\right\rangle\right| \\
&= \tfrac{1}{T}\left|\left\langle (\hat{A}+\hat{C})(\hat{B}+\hat{C})^*\right\rangle\right| \\
&= \tfrac{1}{T}\left|\left\langle \hat{C}\hat{C}^*\right\rangle + \cancel{\left\langle \hat{C}\hat{B}^*\right\rangle} + \cancel{\left\langle \hat{A}\hat{C}^*\right\rangle} + \cancel{\left\langle \hat{A}\hat{B}^*\right\rangle}\right| \\
&= S_c(f)
\end{aligned}
\tag{4}
$$

The Eqn. (4) shows that the cross-terms average to zero and magnitude of the cross-spectrum gives the power spectral density, $S_c(f)$ of the signal $c(t)$. In practice, when the cross-spectrum is calculated, the contribution of $a(t)$ or $b(t)$ is reduced by the square root of the number of averages (N) or the observation time.

***Case ii***: If signal $c(t)$ is presented to the cross-spectrum function in an anti-correlated or phase inverted sense as in Eqn. (5), the magnitude of the averaged cross-spectrum still gives the same result. However, the phase angle of the complex quantity would average to $\pi$.

$$\begin{aligned} x(t) &= a(t)+c(t) \\ y'(t) &= b(t)-c(t) \end{aligned} \tag{5}$$

$$\left|\hat{S}_{xy}(f)\right| = \left|\hat{S}_{xy'}(f)\right| \tag{6}$$

***Case iii***: Now, we introduce a fourth random process, *d*(*t*), another independent, ergodic variable as follows:

$$\begin{aligned} x(t) &= a(t)+c(t)+d(t) \\ y(t) &= b(t)+c(t)+d(t) \end{aligned} . \tag{7}$$

If *d* is also correlated in both *x* and *y,* then the cross-spectrum converges to the sum of the average spectral densities of *c* and *d*.

$$\left|\hat{S}_{xy}(f)\right| \approx S_c(f)+S_d(f) \tag{8}$$

The cross-power spectral density is unable to discern between the two correlated signals and converges to the combination of both. This commonly occurs when amplitude modulation (AM) noise from a source contaminates a phase modulation (PM) measurement via an imperfect mixer [17-19].

***Case iv*** **(Collapse of the cross-spectral function)**: However, if *c*(*t*) is correlated in *x*(*t*) and *y*(*t*) and *d*(*t*) is anti-correlated (phase inverted) in *x* and *y* an unexpected outcome occurs.

$$\begin{aligned} x(t) &= a(t)+c(t)+d(t) \\ y(t) &= b(t)+c(t)-d(t) \end{aligned} \tag{9}$$

The corresponding Fourier transforms and the cross-PSD are represented by

$$\begin{aligned} \hat{X} &= \hat{A}+\hat{C}+\hat{D} \\ \hat{Y} &= \hat{B}+\hat{C}-\hat{D} \end{aligned} \tag{10}$$

$$\begin{aligned} \hat{S}_{xy}(f) &= \tfrac{1}{T}\left\langle \hat{X}\hat{Y}^* \right\rangle \\ &= \tfrac{1}{T}\left\langle (\hat{A}+\hat{C}+\hat{D})(\hat{B}+\hat{A}-\hat{D})^* \right\rangle \\ &= \tfrac{1}{T}\left\langle \hat{C}\hat{C}^* \right\rangle - \left\langle \hat{D}\hat{D}^* \right\rangle + \cancel{\left\langle \hat{A}\hat{B}^* \right\rangle} + \cancel{\left\langle \hat{A}\hat{C}^* \right\rangle} - \cancel{\left\langle \hat{A}\hat{D}^* \right\rangle} + \cancel{\left\langle \hat{C}\hat{B}^* \right\rangle} - \cancel{\left\langle \hat{C}\hat{D}^* \right\rangle} + \cancel{\left\langle \hat{D}\hat{B}^* \right\rangle} + \cancel{\left\langle \hat{D}\hat{C}^* \right\rangle} \end{aligned} \tag{11}$$

$$S_{xy}(f) = \tfrac{1}{T}\left[ \left\langle \hat{C}\hat{C}^*(f) \right\rangle - \left\langle \hat{D}\hat{D}^*(f) \right\rangle \right] \tag{12}$$

What (12) tells us is that at any frequency *f* where the average magnitude of signal *C(f)* is equal to that of signal *D(f)*, the magnitude of the cross-spectrum collapses to zero. Any contribution of the desired signal *c(t)*, or the interferer *d(t)*, to the cross spectral density is eliminated. This occurs even though signals $c(t)$ and $d(t)$ are completely uncorrelated. If the PSD of the two signals are exactly equal, the amount of observed cancelation is limited to $\sqrt{N}$ where $N$ is the number of averages. If *C(f)* and *D(f)* have the same shape or slope versus frequency, entire octaves or decades of spectrum can be suppressed and be grossly under-reported. If the PSD of C and D are not exactly equal a partial cancelation still occurs. The maximum amount of suppression in the magnitude of the cross-spectrum relative to the non-phase inverted measurement (Case iii) approaches

$$\frac{S_c(f)+S_d(f)}{abs(S_c(f)-S_d(f))} \tag{13}$$

at the rate of $1/\sqrt{N}$. Under certain conditions the complex cross-spectrum can actually become negative in value, but since we typically observe the magnitude of the cross-spectrum this is not initially obvious. If the two signals have different frequency dependent noise slopes, a characteristic notch will be observed in the magnitude of the cross spectrum. A differential delay between the two signals presented to the cross-spectrum function will also create a similar notch located at the reciprocal of the delay time.

## Simulation Results: Case iii and iv

Simulations of the collapse of the cross-spectral were created in Matlab Simulink with the block diagram shown in Figure 1.

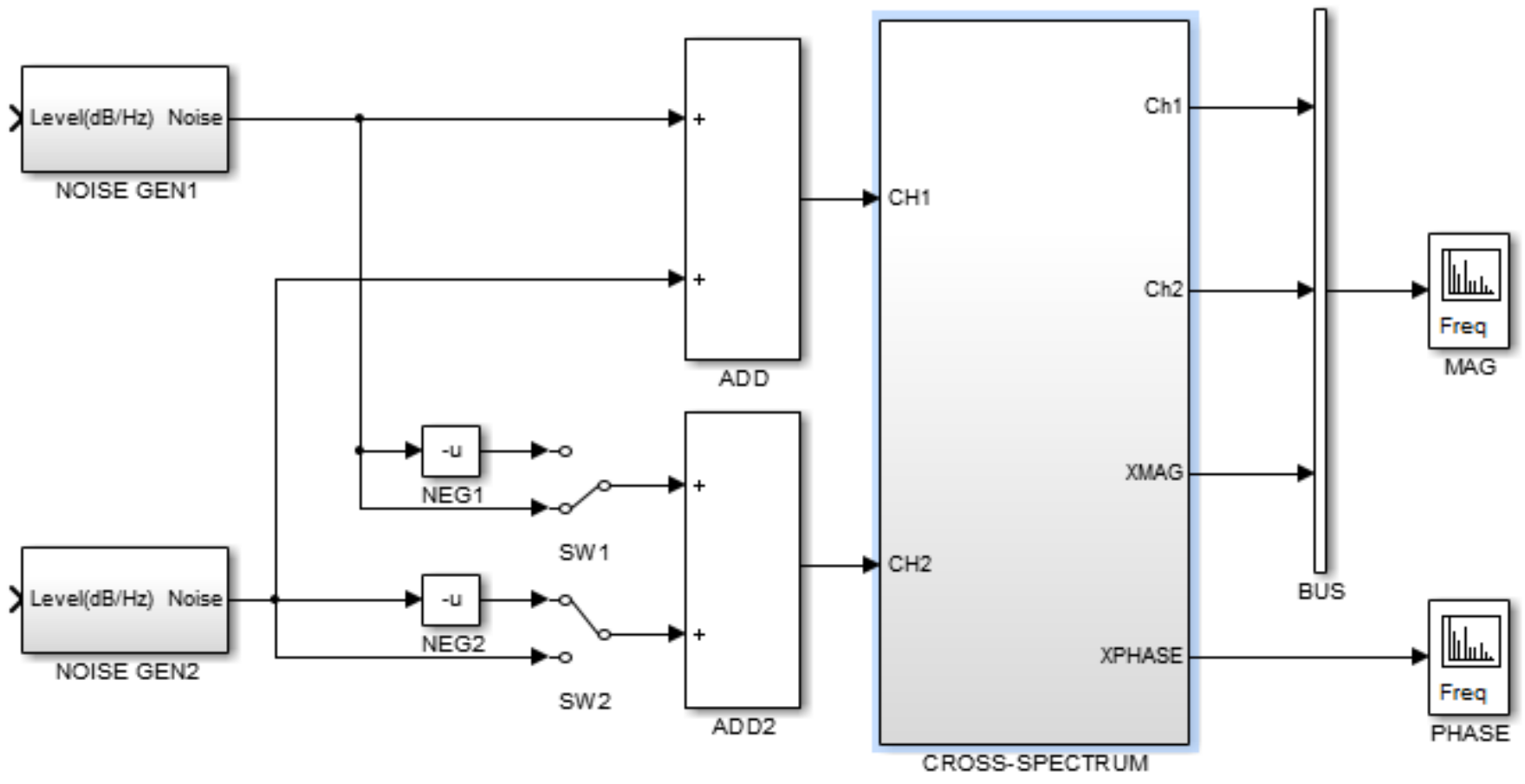

**Figure 1- Block diagram for a Mathworks Simulink simulation. A negation block is switch selectable for creating correlated and anti-correlated inputs to the cross-spectrum.**

Two noise generators, which can create white or frequency dependent noise slopes were summed and connected to both inputs of the cross-spectral density function. Two switches were provided to allow for the negation of either one or both signals to one input of the cross-spectrum. Placing only one or the other switch, but not both, into negation creates the collapse of the function (Figure 2-b). If none or both signals are negated a normal cross-spectrum occurs (Figure 2-a). Finally the collapse due to the interaction of two differently sloped noise types creates a notch is shown in Figure 3.

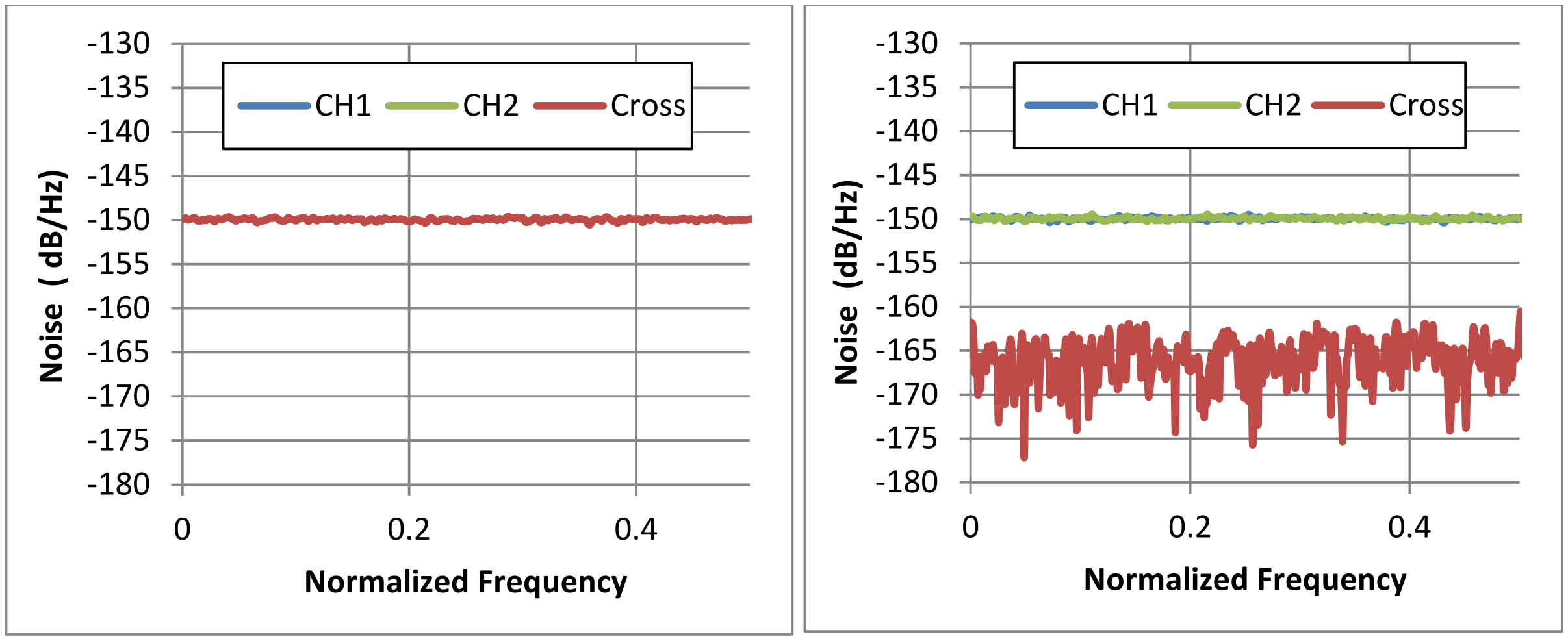

**(a)** **(b)**

**Figure 2- Matlab simulation results for the addition of two completely independent noise sources, *c(t)* and *d(t)*, each with power spectral density of -153 dB/Hz relative to unity. The cross-spectrum for Figure 2-a is when CH1 = CH2 = c*(t)+d(t)* (Case iii). The cross-spectrum for Figure 2-b is when CH1= *c(t)*+d*(t),* and CH2 = c*(t)*-d*(t)* (Case iv). Both figures are for a 1024 point FFT and 1000 averages. The amount of cancelation is 15 dB and follows $\sqrt{N}$.**

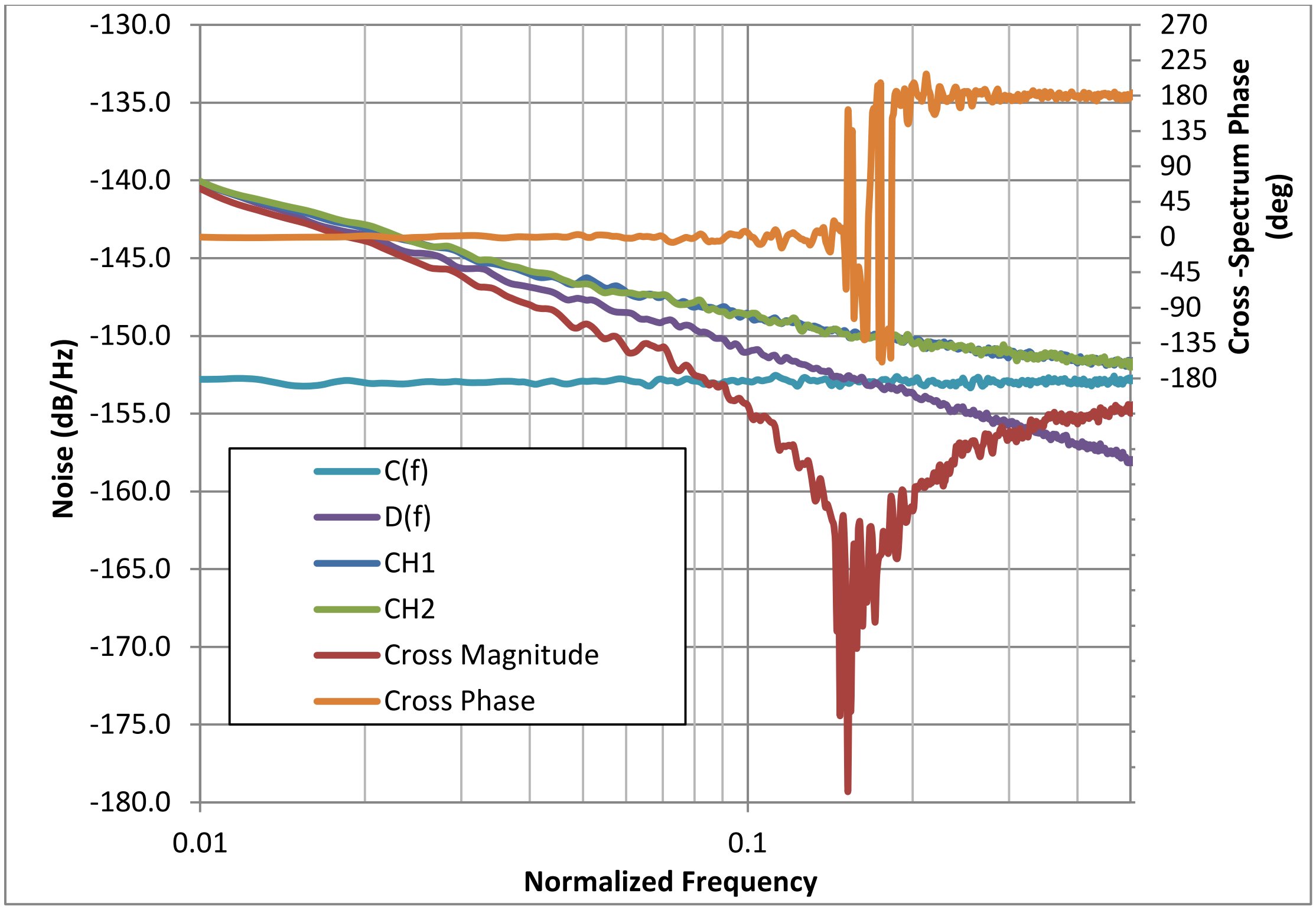

**Figure 3 – Matlab simulation results for the addition of two independent noise sources, c*(t)* and d*(t)*, with different frequency dependence. Signal *C(f)* has a power spectral density of -153 dB/Hz relative to unity. Signal *D(f)* has a $f^{-1}$ slope and intersects signal *C(f)* at a frequency of 0.164 Hz. The cross-spectrum is calculated with CH1= *c(t)*+d*(t),* and CH2 = c*(t)*-d*(t)* (Case iv). Both figures are for a 1024 point FFT and 1000 averages.**

## Cross-spectrum phase noise measurements

A two-channel cross-spectrum phase-noise measurement system for measuring noise properties of an amplifier is shown in Figure 4 [5,15,17]. Each channel includes a phase shifter, a double-balanced mixer as a phase detector and an intermediate frequency (IF) amplifier. The phase shifters establish phase quadrature between two signals at the mixer inputs. AM and PM modulators are also included for evaluating the sensitivities of the measurement system. The output of each mixer, after amplification, is analyzed with a two-channel cross-spectrum fast-Fourier-transform (FFT) spectrum analyzer. The voltage input signals to channel-1 and channel-2 of the FFT analyzer can be written respectively as

$$\begin{aligned} \mathrm{v}_1(t) &= G_1\left[k_{d1}(\Phi_1)\varphi_{DUT}(t) + \beta_1(\Phi_1)\alpha(t) + \mathrm{v}_{B1}(t)\right] \\ \mathrm{v}_2(t) &= G_2\left[k_{d2}(\Phi_2)\varphi_{DUT}(t) + \beta_2(\Phi_2)\alpha(t) + \mathrm{v}_{B2}(t)\right], \end{aligned} \quad (14)$$

where $\mathrm{v}_n(t)$ represents the voltage present at a given FFT channel number ($n$) , $G_n$ is the gain of the baseband intermediate frequency (IF) amplifier, $kd_n$ is the voltage to phase conversion factor of the mixer, $\varphi_{DUT}(t)$ is the phase fluctuations of the device under test (DUT), $\beta_n$ is the AM to voltage conversion factor of the mixer, $\alpha(t)$ is the fractional amplitude fluctuations of the driving oscillator and the DUT and $\mathrm{v}_{Bn}(t)$ is the combined baseband noise of the mixer and IF amplifier. The conversion factors ($k_{dn}$, $\beta_n$) for each mixer are highly dependent on the average phase shift ($\Phi_n$) between its two inputs.

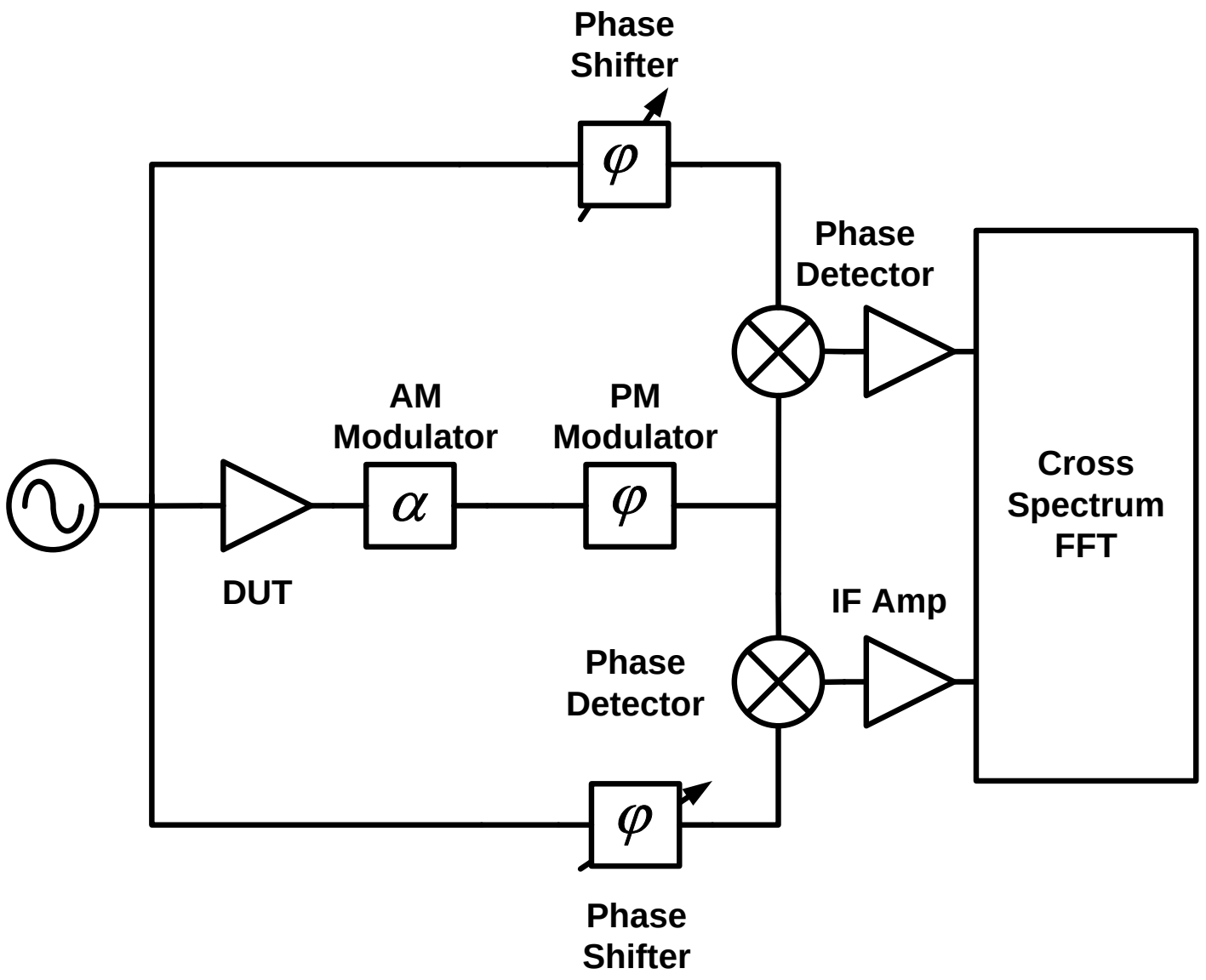

**Figure 4- Block diagram of a two-channel cross-spectrumsystem for measuring PM noise of an amplifier. DUT – Divider Under Test, IF AMP – Intermediate Frequency Amplifier, FFT – Fast Fourier Transform.**

Figure 5 shows the dependence of the conversion factors ($k_d$, $\beta$) on the phase shift between the input signals to an ideal mixer. We can clearly see that that the two sensitivities are orthogonal and one or the other sensitivity is exclusive at the quadrant boundaries (Table 1). Implementing a non-ideal mixer will typically distort the curves, introduce a phase offset and disrupt the orthogonality of the two sensitivities. These aberrations from the ideal double-balanced mixer can be attributed to saturation and non-identically matched diodes and may be highly dependent on operating and environmental conditions. Operating a PM noise measurement with a high level of AM noise rejection can be challenging and without careful tuning and control of the mixer phase shift, $\Phi$, the measurement system will typically show levels of AM rejection around 15 to 30 dB. One would like to operate the mixer in a PM noise measurement at what we call "true quadrature", a phase shift usually near π/2 or 3π/2 where the sensitivity to AM ($\beta$) goes through zero [18, 19]. Finding this exact phase shift typically involves introduction of a sinusoidal AM modulation and tuning the phase shift to minimize the amount of AM leakage into the PM measurement.

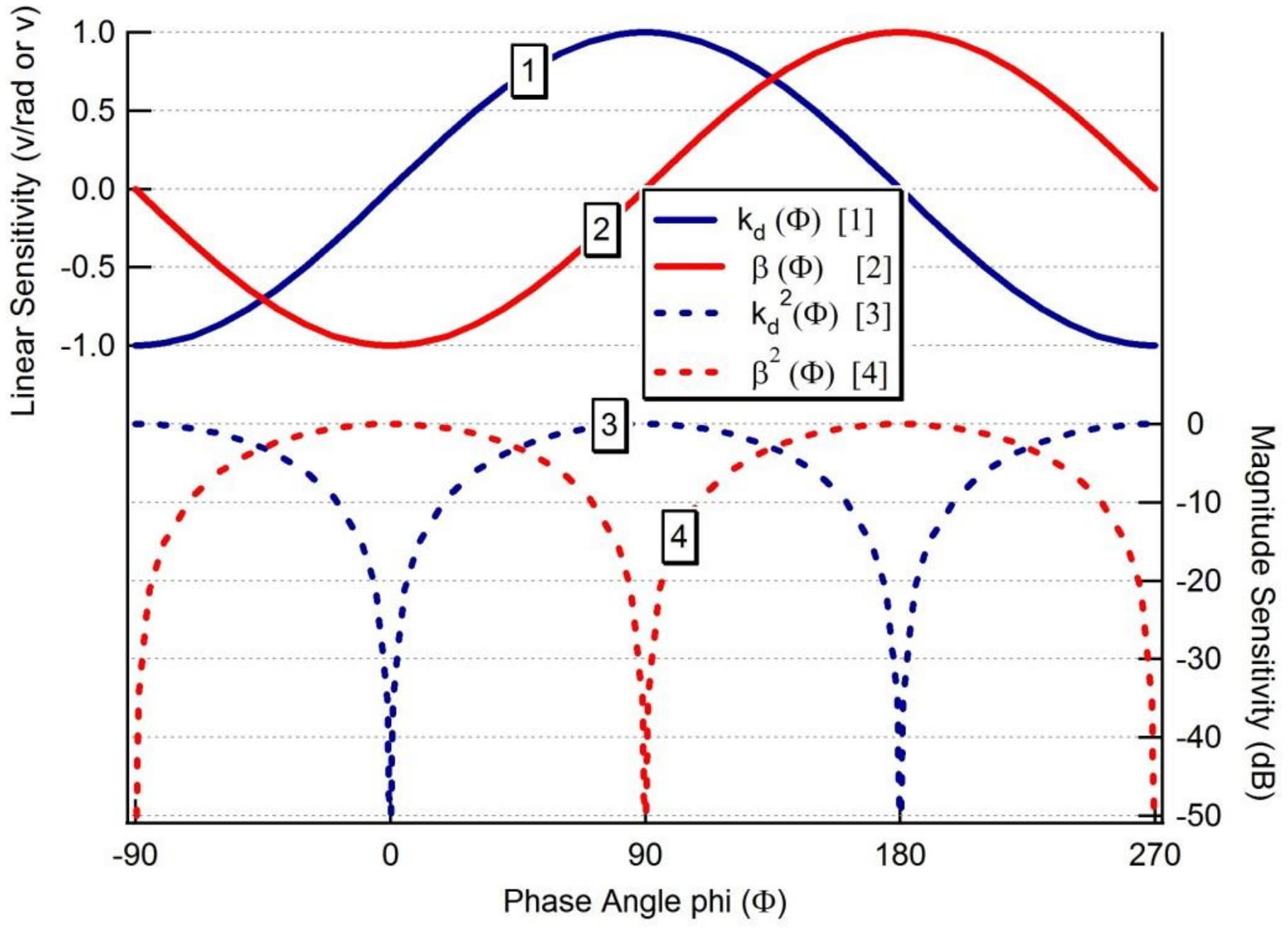

**Figure 5 - Shows the dependence of the conversion factors ($k_d$, β) on average phase shift (Φ) for an ideal mixer operating linearly. The DC output voltage of the mixer will also follow the β(Φ) curve.**

| Quadrant | Range | Sign of $k_d$ | Sign of β |
|---|---|---|---|
| **I** | (0 - π/2) | + | - |
| **II** | (π/2 - π) | + | + |
| **III** | (π - 3π/2) | - | + |
| **IV** | (3π/2 - 2π) | - | - |

**Table 1 – Sign of conversions factors vs. mixer operating quadrant**

Analyzing equation (14), we observe that $\varphi_{DUT}(t)$ is correlated in both $v_1(t)$ and $v_2(t)$ albeit via different conversions factors. The same is true for $\alpha(t)$ while the two baseband signals, $v_{Bn}(t)$ are uncorrelated from each other. Furthermore, the two sets of conversion factors ($k_{d1}$, $\beta_1$) and ($k_{d2}$, $\beta_2$) which dependent on their respective mixer phase shifts, $\Phi_1$ and $\Phi_2$, can be set to 16 different quadrant combinations with varying correlation sense, half which can create correlated/anti-correlated pairs as indicated in Table 2.

| Mixer 2 Quadrant | | Mixer 1 Quadrant | | | |
|---|---|---|---|---|---|
| | | **I** | **II** | **III** | **IV** |
| | **I** | OK | X | OK | X |
| | **II** | X | OK | X | OK |
| | **III** | OK | X | OK | X |
| | **IV** | X | OK | X | OK |

**Table 2- Mixer quadrants pairs that may cause undesired cross-spectral cancelation. 'OK' indicates that each conversion factor has the same sign of correlation in both mixers. 'X' indicates a correlated and anti-correlated (Case iv) pair of conversions factors ($k_d$, β).**

When high levels of AM rejection are required and thus operating near the AM null, environmental and signal power changes may cause the location of the null or operating point to drift causing the sign of $\beta$ to flip. This may cause the noise cancelation effect to appear and disappear temporally, further masking its presence during an average. Measuring low levels of PM noise in the presence of much higher AM levels requires continuous operation near the bottom of the null and may require pure AM modulators [20, 21] and an active control servo to maintain a high level of AM suppression.

## Experimental results

In order to experimentally demonstrate the collapse of the cross-power spectral function the setup described in Figure 4 is used. An amplifier with high gain and low input power is used as the DUT to generate a signal with a high white PM noise level. The DUT has 42 dB of gain, a noise figure of 7 dB and is operating with an input power of -25 dBm creating a white PM noise level of L(>10 kHz) = -145 dBc/Hz. Under these conditions, the noise of DUT dominates and the mixer's residual noises are neglectable. The PM noise of the DUT is shown as trace (4) in Figure 6. An AM modulator is driven with white noise from a direct digital synthesizer (DDS) producing an AM level of -130 dBc/Hz shown as trace (2). The phase shifters of both mixers are detuned from "true quadrature" so that the relative level of AM suppression ($k_d/\beta$) of 15 dB is achieved. The amount of AM leakage into the PM noise measurement is measured and shown as trace (5). If we perform a cross-spectrum PM noise measurement with both mixers in the same quadrant (Case iii), we observe the expected result shown in trace (3). In this case the measured white PM noise is twice that of the DUT due to the presence of an equal amount of leakage white AM noise. If we now adjust the mixers to have the same amount of AM leakage, but place them in different quadrants (**I** and **II**) so that that $K_{d1} = K_{d2}$ and $\beta_1 = -\beta_2$, a correlated/anti-correlated pair of conversion factors are created (Case iv) and we now observe the cross-spectrum approaches zero in the white region as shown in trace (6). This annihilation occurs in the cross-spectrum even though the white PM noise of the amplifier and the leaked injected AM noise are entirely uncorrelated. The phase of the cross-spectrum, trace (1), shows that in the region where the cross-spectrum is still average limited, the phase is random.

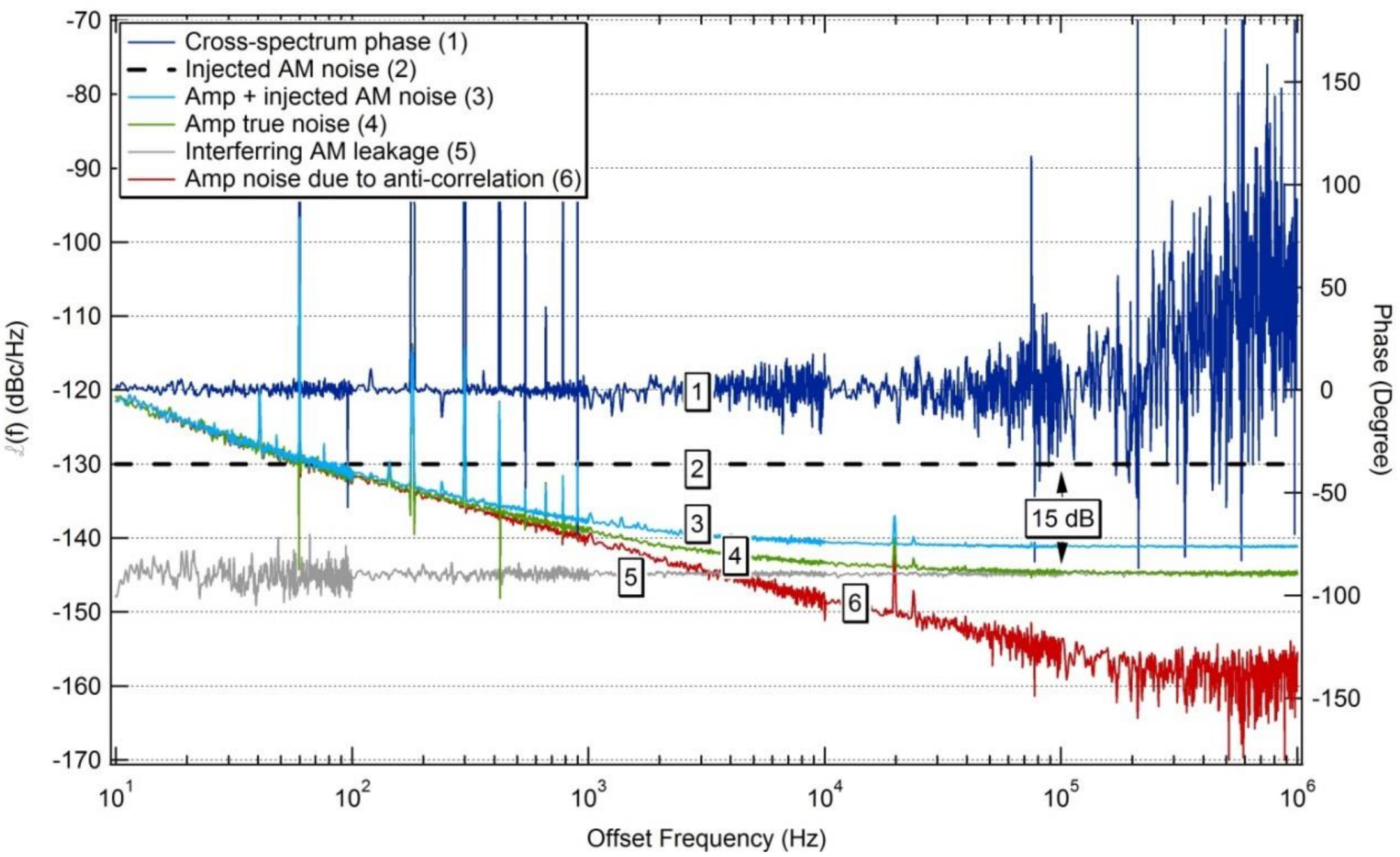

**Figure 6 – Experimental results demonstrating the annihilation of DUT PM noise in a cross-spectrum measurement due to AM leakage.**

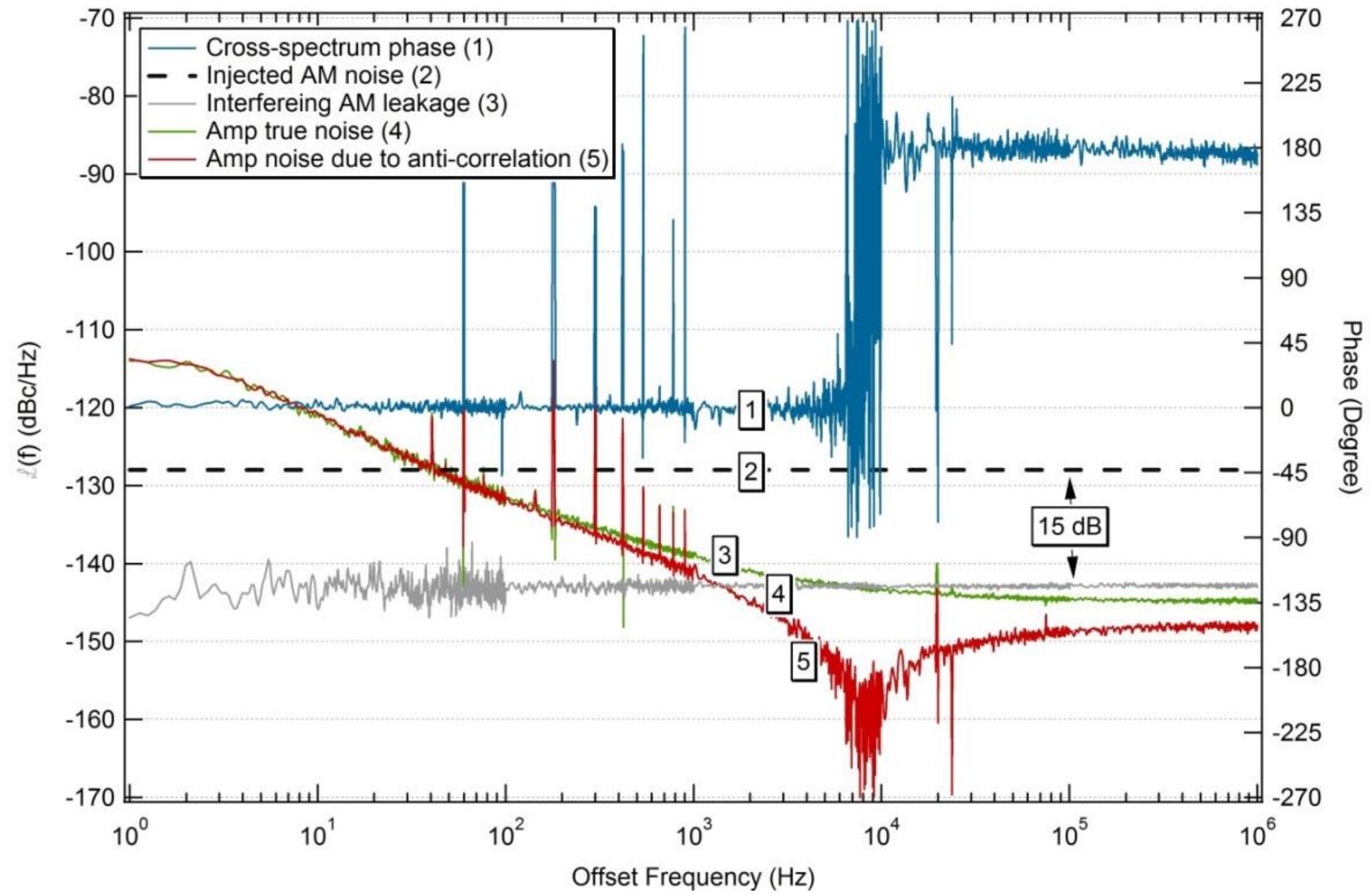

**Figure 7 - Experimental results demonstrating the notch that occurs when two noise processes with different slopes intersect while one is correlated and the other is anti-correlated in the cross-spectrum**

Another example is demonstrated when we reduced the amount of AM rejection in the mixer so that the level of leakage AM noise is slightly higher than the thermal noise of the amplifier. This causes the two noise signals to cross at an offset frequency around 10 kHz creating a notch and grossly underestimated the noise from 300 Hz and above. The phase of the cross-spectrum in Figure 7 clearly shows that the AM noise which, dominant above 10 kHz, is anti-correlated with a phase angle of π, and the PM noise, dominant below 10 kHz, has an angle of zero. In the region where the cross-spectrum is still average limited, the phase is random.

The cancelation of noise is not just limited to AM and PM noise interactions. Any two noise sources that present themselves to both inputs of the cross-spectrum function as correlated and anti-correlated can cause this effect. A final experimental case was found where AC power line noise was coupling into the IF amplifier noise floors and interacting with an anti-correlated phase noise from the DUT. The coupled power supply noise was at a similar level as the DUT noise producing a notch as indicated in Figure 8.

The correlation null has also been observed in the datasheets of commercial devices. A particular example is shown in Figure 9 where $f^{1}$ AM noise leakage interacts with the $f^{3}$ and thermal noise corner.

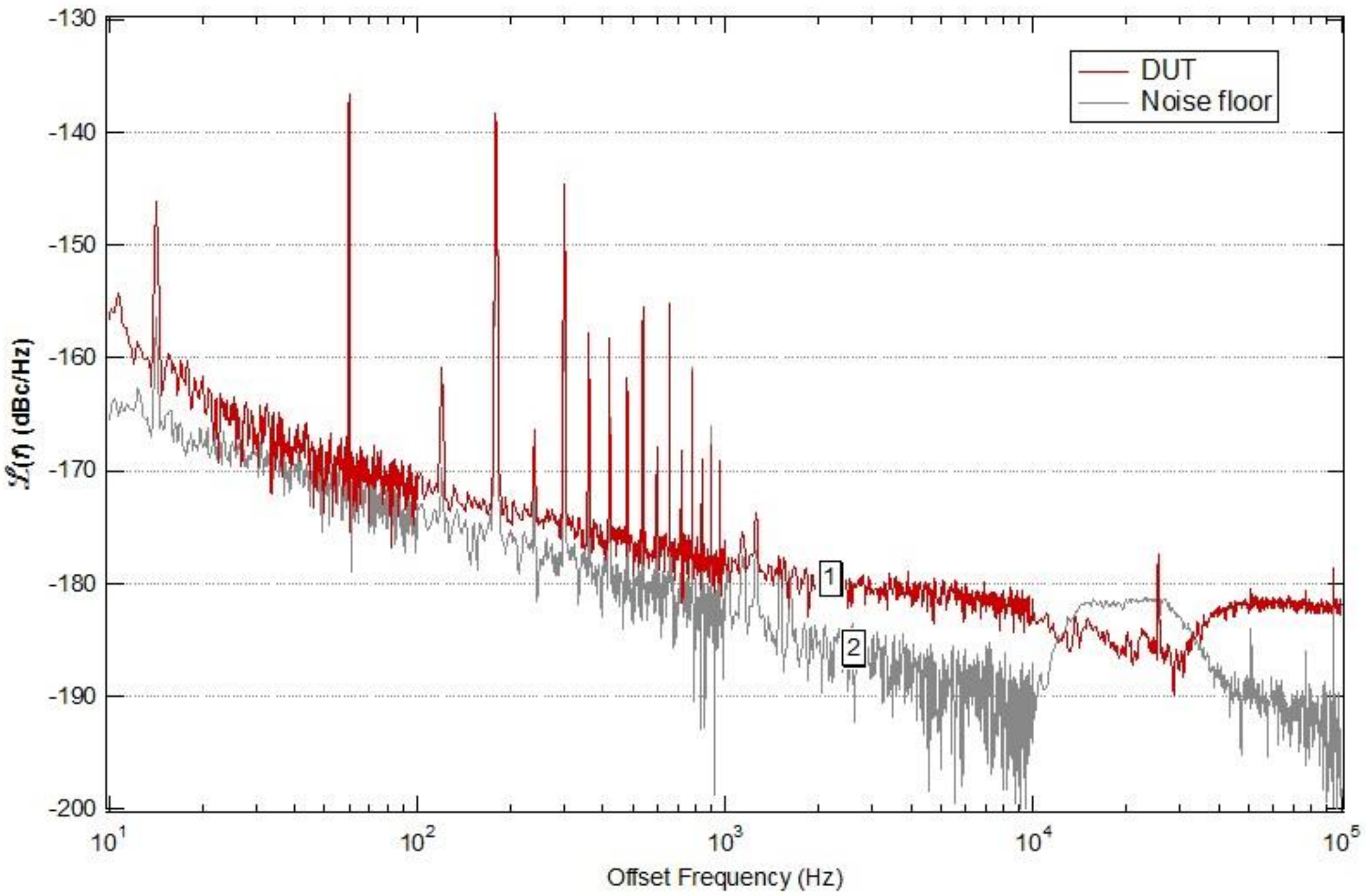

**Figure 8 – Cross-spectral null observed when correlated/anti-correlated power supply noise interacts with an amplifier measurement.**

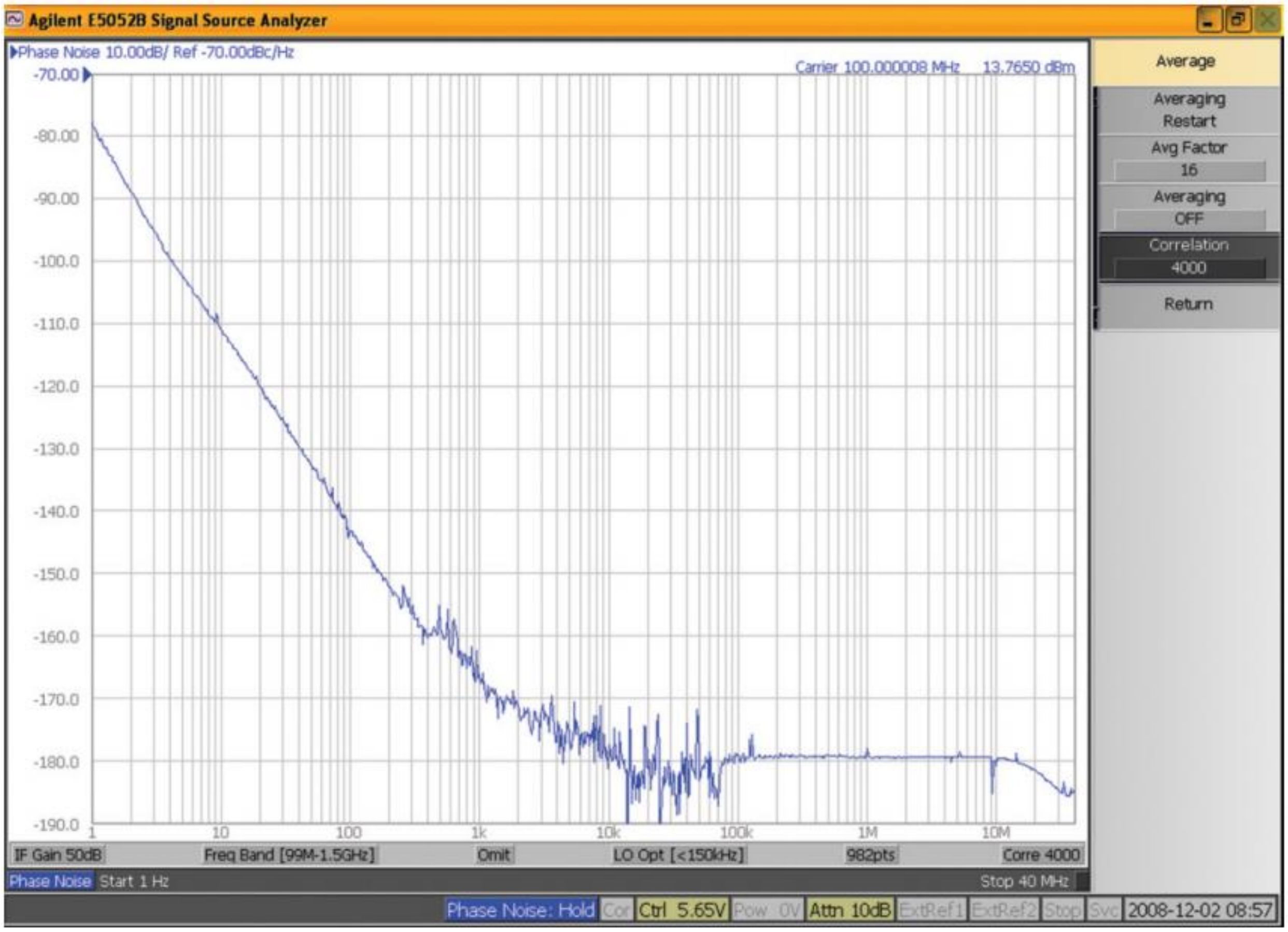

**Figure 9 – Observance of a cross-spectral null from the datasheet of a commercial oscillator measured with an Agilent cross-spectrum phase noise test set.**

## Mitigation of the collapse in cross-spectral phase noise measurements

The detection of the cross-spectral collapse after a measurement can be difficult. If an interfering signal has a phase inversion relative to the desired signal in one channel of the cross-spectrum, and has an average magnitude within +/- 10 dB of the desired signal, a collapse of at least 1dB will be observed. This can be as high as $\sqrt{N}$ if the magnitudes are closely matched. Observing the phase of the cross-spectrum of the final measurement data can be of little help. If the measurement of the cross-spectrum is still averaged limited, its phase will be random. When the interfering signal is higher than the desired signal in a region of the spectrum as in Figure 7, a phase transition will be observed in addition to the notch in the magnitude. If the desired signal and the interfering signal have the same frequency dependent slope, it can be difficult to determine whether or not a collapse has occurred. The best method of mitigation is to analyze the interfering signals and the conversion factors prior to a measurement with the following steps:

1. Measure the AM noise of the source.
2. Adjust mixer phase shift to ensure AM leakage is sufficiently low for both mixers.

Use an AM modulator to inject an AM tone into the PM measurement and adjust the mixer quadrature point ($\Phi$) for minimum leakage for both mixers. Realize that the point of minimum AM leakage is also point of highest sensitivity, small changes in phase can lead to large changes in magnitude as well as the sign of the $\beta$ conversion factor. Environmental and operating point changes can also affect $\beta$. Using a non-pure AM modulator can lead to an underestimation of the AM suppression factor and an incorrect determination of the location of "true quadrature". If high levels of AM rejection are required, pure AM modulators [18,19] and an active servo control of the quadrature phase shift may be a necessity. If active control is not employed, it may be beneficial to sacrifice some suppression by operating at an offset from the minimum and ensuring that changes in the sign of $\beta$ will not occur due to operating point fluctuations.

3. Measure the amount AM-to-PM leakage and ensure that the AM source noise leakage is at least 10 dB below the expected PM level.
4. Ensure that both mixers are in the same operating quadrant

   Measure the amplitude and phase conversion factors for both mixers and confirm that both mixers are operating in the same quadrant. It is important that the conversion factors for any signals presented to the cross-spectrum have the same sign in both inputs. Tuning the phase shift and observing the direction of change in the DC output voltage of the mixers can determine if the mixer is in quadrant set (I, II) or (III, IV). Using AM and PM modulators and detecting the cross-spectral phase of the modulated tones can also determine if the mixers are creating phase inverted pairs or not. This can be a powerful tool for setting up a correct cross-spectral analysis.

5. Change the quadrant for one mixer so that $\beta$ changes sign but has the same absolute value while $k_d$ remains the same, and observe the change in the cross-spectrum. If there is a dramatic change, one of the two quadrants may be creating a collapse.
6. NEVER use a low measured noise level as an indicator for finding "true quadrature."

The discussion in this paper focuses on the case where the desired and interfering signal are independent, however it is also important to consider that for multiplicative, or non-thermal noise types, AM and PM noise may be correlated even if it has non-identical frequency dependent slopes [22] and may cause an entirely different set of problems when using a correlator.

## Discussion and Conclusion

We demonstrate and explain a set of conditions where the detection of a desired signal in cross-spectral analysis fails partially or entirely. If two time series, each comprised of the summation of two fully independent signals, are correlated in the first time signal and anti-correlated (phase inverted) in the second, and have the same average spectral magnitude, the cross-spectrum power density between two time series collapses to zero. These two conditions may occur only at localized offset frequencies or over a wide range of frequency of the cross-spectrum. The anti-correlation of one of the signals relative to the other may be caused by phase inversion, negation, time-delay or any other mechanism. This condition initially seems counter-intuitive because the introduction of a new, independent noise signal typically does not cause a reduction of total measured noise. The simultaneous presence of correlated and anti-correlated signals can lead to gross underestimation of the total signal in cross-spectral analysis. Partial collapse of the functions still occurs when the spectral densities are close in magnitude but not equal. Cross-spectral PM noise measurements suffering from AM leakage can be particularly sensitive to these underestimations because adjustment of the mixer phase shift can easily generate correlation/anti-correlation pairs and the level of AM leakage is widely variable and dependent on phase shift. Also, cross-spectrum analysis is commonly used in all-digital measurement systems [13] to reject the relatively high level of uncorrelated quantization noise in the analog-to-digital converters by 20-30 dB. This same quantization noise can also produce AM-to-PM conversion in the digital quadrature/in-phase detectors. We believe that cross-spectrum collapse may occur less frequently in digital systems because the level of AM leakage is not as prominent and more predictable in heterodyne digital down-converters. However, cancelation notches have still been observed in digital measurement systems after long averaging times. Finally, the photonic microwave generation from optical stabilized sources can lead to extremely low PM noise in the presence of much higher AM noise [21,22]. Under these conditions the danger of collapse is high and cross-spectrum must be evaluated and used very carefully.

## Acknowledgements

The authors thank Neil Ashby, Franklyn Quinlan, Tara Fortier, Scott Diddams, Fred Walls, and Eugene Ivanov for theoretical and experimental discussions. We also thank Danielle Litrette, David Smith, Mike Lombardi and for help with preparation and editing this work.